\documentclass[12pt, a4paper]{article}

\pdfoutput=1

\usepackage{amsmath}
\usepackage{amsfonts}
\usepackage{amssymb}
\usepackage{bbm}
\usepackage{verbatim}
\usepackage{dsfont}
\usepackage{booktabs}
\usepackage{slashed}

\usepackage{epsfig}
\usepackage{color}
\usepackage[table]{xcolor}
\usepackage{graphicx}
\usepackage[]{caption}
\usepackage[listofformat=empty,subrefformat=empty]{subfig}
\usepackage{float}
\usepackage{overpic}

\usepackage{enumerate}
\usepackage{hhline}
\usepackage{multirow}

\usepackage{cite}
\usepackage{xspace}
\usepackage{setspace}

\usepackage[pdftitle={de Sitter vacua from an anomalous gauge symmetry},
  pdfauthor={Wilfried Buchmuller, Markus Dierigl, Fabian Ruehle, Julian Schweizer},
  pdfsubject={de Sitter, D-term uplift, Supergravity, Axions},
  bookmarksopen, bookmarksnumbered, bookmarksopenlevel=2, colorlinks=false, linkcolor=blue, citecolor=blue, urlcolor=blue]{hyperref}

\renewcommand{\tfrac}{\genfrac{}{}{}1}

\begin{document}

\thispagestyle{empty}

\begin{flushright}
DESY-16-030\\
\end{flushright}
\vskip .8 cm
\begin{center}
{\Large {\bf de Sitter vacua from\\[2pt] an anomalous gauge symmetry
}}\\[12pt]

\bigskip
\bigskip 
{
{\bf{Wilfried Buchmuller}\footnote{E-mail: wilfried.buchmueller@desy.de}},
{\bf{Markus Dierigl}\footnote{E-mail: markus.dierigl@desy.de}},  
{\bf{Fabian Ruehle}\footnote{E-mail: fabian.ruehle@desy.de}},
{\bf{Julian~Schweizer}\footnote{E-mail: julian.schweizer@desy.de}}
\bigskip}\\[0pt]
\vspace{0.23cm}
{\it Deutsches Elektronen-Synchrotron DESY, 22607 Hamburg, Germany }\\[20pt] 
\bigskip
\end{center}

\begin{abstract}
\noindent
We find a new class of metastable de Sitter solutions in compactifications of
six-dimensional supergravity motivated by type IIB or heterotic string vacua. Two Fayet-Iliopoulos terms of a
local $U(1)$ symmetry are generated by magnetic flux
and by the Green-Schwarz term canceling the gauge anomalies, respectively.
The interplay between the induced D-term, the moduli dependence of the effective gauge coupling, and a nonperturbative superpotential stabilizes the moduli and determines the size of the extra dimensions.
\end{abstract}

\newpage 
\setcounter{page}{2}
\setcounter{footnote}{0}

{\renewcommand{\baselinestretch}{1.3}

\section{Introduction}
\label{sec:Introduction}

The observed accelerated expansion of the universe \cite{Perlmutter:1998np} is most
easily explained by a constant vacuum energy density, the characteristic
feature of de Sitter space. This has led to an intense search for de
Sitter vacua in supergravity and superstring theories, which represent
attractive extensions of the standard model of particle
physics. Although much progress has been made, challenging questions 
still remain \cite{Douglas:2006es}.

A particularly attractive proposal by Kachru, Kallosh,
Linde and Trivedi \cite{Kachru:2003aw} is based on quantized fluxes in
type IIB string theory and additional nonperturbative effects
stabilizing the overall volume modulus. These features are captured in
the KKLT superpotential that yields an anti-de Sitter minimum. To achieve
the required `uplift' to de Sitter space additional ingredients are
needed such as anti-D3 branes \cite{Kachru:2003aw,Choi:2005ge},  $D$-terms induced by magnetic
flux \cite{Burgess:2003ic, Dudas:2005vv, Braun:2006se},
a gauged R-symmetry \cite{Villadoro:2005yq, Antoniadis:2014hfa} or T-branes \cite{Cicoli:2015ylx}.
Alternatively, additional matter fields have to be added
allowing for an F-term uplift (see e.g. \cite{Lebedev:2006qq, Kallosh:2006dv}). Also
extensions of the KKLT proposal, such as the Large Volume Scenario 
\cite{Balasubramanian:2005zx} or K\"ahler Uplift
\cite{Balasubramanian:2004uy, Westphal:2006tn} need additional 
ingredients to achieve an uplift to de Sitter vacua.

Conditions for  de Sitter vacua in four-dimensional supergravity theories
derived from string theory compactifications have previously been
analyzed in \cite{Covi:2008ea, Covi:2008zu}. In particular,  
constraints on the K\"ahler potential and, in case of gauged shift
symmetries, on the associated Fayet-Iliopoulos (FI) terms have been derived \cite{GomezReino:2007qi}.
For a related discussion of de Sitter and Minkowski vacua in string theory
see, for instance, \cite{Blumenhagen:2003vr,Dudas:2005vv, Villadoro:2006ia, Cicoli:2013rwa, Lukas:2015kca, Retolaza:2015nvh}.

In this Letter, we study de Sitter vacua in six-dimensional
supergravity models, compactified on orbifolds with flux. Such models
have been suggested as an intermediate step in string
compactifications to four dimensions \cite{Hebecker:2004ce}. Our results
represent an application of previous work \cite{Buchmuller:2015eya, bdrs3}. 
The crucial ingredients are the flux and the Green-Schwarz term of an
anomalous local $U(1)$ symmetry. They lead to two Fayet-Iliopoulos
terms of opposite sign, and they modify the gauge kinetic function.
Together with a nonperturbative KKLT-type superpotential one naturally
obtains de Sitter vacua, without any additional degrees of freedom below the flux induced mass scale. 
In the following we derive relations between the superpotential
parameters and de Sitter vacua in moduli space and give an
explicit example.

\section{Conditions for de Sitter vacua}
\label{sec:model}

In this work we consider a simple $\mathcal{N}=1$ supergravity model\footnote{We follow the conventions of \cite{Wess:1992cp}.} that has been derived as the effective four-dimensional Lagrangian for flux compactifications of six-dimensional supergravity \cite{Buchmuller:2015eya, bdrs3}. It contains three chiral multiplets, $S$, $T$, $U$ and the real
vector multiplet $V$ of a $U(1)$ gauge symmetry. The K\"ahler potential is given by 
\begin{align}
K = - \log (S + \overline{S} + i X^S V) - \log (T + \overline{T} + i X^T V) - \log (U + \overline{U}) \,.
\label{kahler}
\end{align}
where $X^{S,T}$ are purely imaginary and constant Killing
vectors parameterizing a gauged shift symmetry. Note that the
K\"ahler potential is of no-scale type, i.e.\ 
$K_i K^{i \bar{\jmath}} K_{\bar{\jmath}} = 3$,
where $i$ ($\bar{\jmath}$) denotes the derivative with respect to the $i$-th
chiral ($\bar{\jmath}$-th anti-chiral) multiplet and $K^{i\bar{\jmath}}$ is the inverse
K\"ahler metric.

To completely define the gauge sector one further has to specify the gauge
kinetic function $H$, whose real part $h$ corresponds to the effective
gauge coupling. In the following it will be crucial that, in addition to the
classical term linear in $S$, the gauge kinetic function contains a second part linear in $T$,
\begin{align}
H = h_S S + h_T T \,.
\end{align}
A $T$-dependence is known to arise due to quantum corrections
\cite{Ibanez:1986xy,Dixon:1990pc}. Following the convention $S = \tfrac{1}{2} (s + i c)$, $T = \tfrac{1}{2} (t + i b)$, one has
$h= \tfrac{1}{2} (h_S s + h_T t)$.
Finally, we have to specify the superpotential. It has to be gauge
invariant and can therefore only depend on the linear combination 
\begin{align}
Z = -i X^T S + i X^S T \equiv \tfrac{1}{2} (z + i \tilde{c})\,,
\label{zmoduli}
\end{align}
where in a first step, we ignore the shape modulus $U$.
For reasons discussed above we consider a KKLT-type potential,
\begin{align}\label{super}
W (Z) = W_0 + W_1 e^{- a Z} \,,
\end{align}
where, w.l.o.g.\ we choose $a>0$, $W_0$ and $W_1$ to be real. The third modulus $U$ can be stabilized by inclusion of a term $W_2 \exp{(-a'U)}$ in the superpotential. However, the resulting equations are considerably more involved than in the two moduli case. They will be given in \cite{bdrs3}.

In the following it will be important that the parameter $h_T$ in
the gauge kinetic function is negative while the classical contribution
$h_S$ is positive. Furthermore,
the two Killing vectors have to be of opposite sign. As we shall see
in the next section, both conditions are satisfied 
in flux compactifications of 6d supergravity.
The line along which $h(s,t) = 0$ divides the moduli space into a
physical ($h > 0$) and an unphysical region ($h < 0$). The physical
region thus satisfies the condition
\begin{align} 
t < t_{(h)} =-\frac{h_S}{h_T} s \,.
\label{physicalregime}
\end{align}
The scalar potential is a sum of $F$- and $D$-term contributions,
\begin{align}\label{pot}
V = V_F + V_D = e^{K}(K^{i \bar{\jmath}} D_i W D_{\bar{\jmath}} \overline{W} - 3 |W|^2) + \frac{g^2}{2 h} D^2 \,,
\end{align} 
where $D_i W = W_i + K_i W$ is the
K\"ahler covariant derivative of the superpotential and $g$ is a gauge coupling, resulting in $g/\sqrt{h}$ as the effective gauge coupling. The $D$-term is given in terms of
the Killing vectors,
\begin{align}
D = i K_i X^i = -\frac{i}{s} X^S - \frac{i}{t} X^T \,.
\label{Dterm}
\end{align}
Due to their opposite sign $D=0$ defines a second line in $(s,t)$
space given by
\begin{align}
t_{(D)} = - \frac{X^T}{X^S} s\,.
\label{Dzero}
\end{align}
If this line is not part of the physical region in moduli space, i.e.\
$|X^T/X^S| \geq |h_S/h_T|$, the $D$-term potential is positive definite.
This is the situation we want to study in the following.

Solutions of the equations of motion leading to Minkowski or de Sitter vacua have to fulfill,
\begin{align}\label{extrema}
\partial_S V =0\ ,\quad \partial_T V = 0\ ,\quad V = \epsilon \geq 0\,,
\end{align}
which yield three relations between the $F$-term and $D$-term
contributions to the potential.
It is convenient to use 
the linear combinations $\partial_{\pm} = s\partial_S
\pm t \partial_T$ instead of $\partial_{S,T}$ . 
Derivatives of the superpotential then produce the
factors $\partial_- Z = stD$ and $\partial_+ Z = stE$, where
\begin{align}
E = i K_T X^T - i K_S X^S = -\frac{i}{t} X^T + \frac{i}{s} X^S \,.
\label{Eparameter}
\end{align}
In terms of $D$ and $E$ the scalar potential can be written in the compact form
\begin{align}\label{simple}
V = \frac{s t}{2} (D^2 + E^2) A - E B + \frac{g^2}{2 h} D^2 \,,
\end{align}
with
\begin{align}
A &= |\partial_Z W|^2 = a^2 \, W_1^2 \, e^{-a z} \,,\label{W1}\\
B &=2 \text{Re}\left(\partial_Z W \overline{W}\right)
= -2 a W_1 \left(W_1 \, e^{-a z} + W_0
  \, e^{-\tfrac{a}{2} z} \cos\left(\tfrac{a}{2} \, \tilde{c}\right)\right)\,.
\label{W0}
\end{align}
Note that the term $-3|W|^2$ in the scalar potential \eqref{pot} has disappeared due to the
no-scale structure of the K\"ahler potential. Choosing opposite signs
for $W_0$ and $W_1$, the axion $\tilde{c}$ is stabilized at zero, and in
the following we set the cosine term to one. 
The orthogonal axion gives a mass to the $U(1)$
vector field via the Stueckelberg mechanism \cite{Buchmuller:2015eya}.

In order to find minima we invert the problem and solve for the superpotential parameters in terms of $s$ and $t$. In this way we can see which superpotential parameters lead to minima in the $(s,t)$ plane for realistic parameter ranges.
The conditions for extrema \eqref{extrema} yield three relations
between the superpotential parameters and the position $(s,t)$ of the extrema in moduli space. A straightforward calculation leads to
\begin{equation}\label{Ast}
\begin{aligned}
A &= - \frac{1}{2 h^2 st (1 - \rho^2)} \left( h_T t \rho + h (2 - \rho + \rho^2) +
h^2 \frac{2 \epsilon}{E^2} \right) \,, \\
B &= - \frac{E}{4 h^2 (1 -
  \rho^2)} \left(h_T t \rho (1 + \rho^2) + h (2 - \rho + \rho^2 -
  \rho^3 + 3 \rho^4)  + h^2 \frac{8 \epsilon}{E^2}   \right)\,, \\
a&= -\frac{2 E (1 - \rho^2)}{st} \enspace \frac{h_T t \rho + h (2 - \rho - 3 \rho^2)}{E^2 \left( h_T t \rho (1 + \rho^2) + h (2- \rho + 5 \rho^2 - \rho^3 - \rho^4)\right)+8 h^2 \rho^2 \epsilon} \,, 
\end{aligned}
\end{equation}
where we have introduced the ratio $\rho = D/E \in (-1, 1)$. The
quantities on the r.h.s. of these equations, $h$, $E$ and $\rho$, only
depend on the values of the moduli fields $s$ and $t$,  whereas $A$,
$B$ and $a$ are related to the superpotential by the definitions
\eqref{W1} and \eqref{W0}.

Eqs.~\eqref{Ast} can now be used to
identify consistency conditions for the existence of de Sitter vacua
and to construct explicit solutions.
A similar discussion for pure $F$-term breaking has recently been carried
out in \cite{Kallosh:2014oja}. From Eq.\ \eqref{W1} we see that $A$
has to be positive, and by definition $a > 0$. Furthermore, $B>0$ is
needed for a cancellation of the positive contributions in the
potential \eqref{simple}, thus allowing for a vanishing or small
cosmological constant. From Eqs.~\eqref{Ast} one reads off that these conditions can be satisfied if $h_T < 0$. The reason is that the terms proportional to $h$, which would
give the unwanted sign, can be made small due to the
opposite sign of the two contributions $h_S s$ and $h_T t$. Given a
solution for a chosen extremum at $(s,t)$, one then has to examine
whether the obtained values for $h$, $a$, $W_0$ and $W_1$ are
physically meaningful.

The de Sitter solutions obtained this way correspond to metastable vacua since the familiar runaway solution at infinity has zero energy density. Due to the similarity with the KKLT model one expects no further vacua. We have checked this numerically for various sets of superpotential parameters and we have indeed found no other minima. Let us finally emphasize again that the obtained Minkowski and de Sitter minima are
a consequence of the opposite signs of the moduli contributions to 
both the gauge kinetic function and the $D$-term in combination with the nonperturbative superpotential.

In the de Sitter vacua constructed this way supersymmetry is broken by a
$D$-term since,  as discussed above, $D > 0$ in the physical region of the moduli
space. Vanishing F-terms, i.e.\ $D_S W = D_T W = 0$, would imply
$D\partial_ZW = 0$, which is not possible for $D > 0$ and a KKLT-type
superpotential. Hence, as expected, supersymmetry is also broken by
$F$-terms and the Goldstino is a mixture of the gaugino and modulini, with
$m_{3/2} \sim \langle F\rangle \sim \langle D\rangle$.
It is well known that no-scale models require carefully chosen superpotentials in order to allow for metastable de Sitter vacua \cite{Covi:2008ea, Covi:2008zu}. In the above analysis simple solutions are found due to the gauged shift symmetry, a possibility which has already been discussed in \cite{GomezReino:2007qi}.

\section{An example}
\label{sec:example}

In \cite{bdrs3} we derive the low-energy effective action for a
six-dimensional supergravity model with a $U(1)$ gauge field and a
charged bulk matter field, compactified on an orbifold
$T^2/\mathbb{Z}_2$ with magnetic flux. The
antisymmetric tensor field couples to the $U(1)$ gauge field via a
Chern-Simons term. The four-dimensional effective action 
involves the K\"ahler potential \eqref{kahler}
with a gauged shift symmetry in $S$ and $T$. The Killing vectors are given by
\begin{align}\label{XTS}
X^T = -i\frac{f}{\ell^2}\,, \quad X^S = -i g^2 \frac{N+1}{(2\pi)^2}\,,
\end{align}
where $f$ is the quantized flux, $f = -4\pi N < 0$, $N \in\mathbbm{N}$,
and $\ell$ is a length scale in Planck units. $X^S$ is a consequence
of the Green-Schwarz term which is needed to cancel bulk and fixed
point anomalies. It represents a quantum correction which, to our knowledge,
has been neglected in previous analyses of moduli stabilization. Note that the two Killing
vectors contribute to the $D$-term with opposite signs: $-iX^T > 0$, $-iX^S < 0$.

The orbifold compactification with $N$ flux quanta generates $N+1$
massless Weyl fermions, as discussed in \cite{Buchmuller:2015eya}. Their scalar superpartners $\phi_i$ obtain masses from
the $D$-term potential
\begin{align}
V_D = \frac{g^2}{2h}\left(\sum_{i=1}^{N+1} |\phi_i|^2 + D\right)^2\,,
\end{align}
with $D$ given by Eq.~\eqref{Dterm}. The stabilization
of the charged matter fields $\phi_i$ at the origin requires $D>0$, as assumed in
the previous section. 
Note that all matter fields $\phi_i$ have the same $U(1)$ charge
since they originate from the same bulk field.
The size of $D$ determines the mass
splitting within matter multiplets and therefore the scale of
supersymmetry breaking \cite{Bachas:1995ik}. Furthermore, we want to point out that the quantity $E$ introduced above is
always positive, $E = iX^S/s -iX^T/t > 0$, and $\rho = D/E \in (0, 1)$.

As discussed in the previous section, the existence of de Sitter
minima crucially depends on the gauge kinetic function for which
one obtains
\cite{bdrs3}
\begin{align}\label{hST}
h_S = 2\, \quad h_T = -\frac{2g^2\ell^2}{(2\pi)^3}\,.
\end{align}
At the classical level, the gauge kinetic function only depends on
$S$. The contribution proportional to $T$ is a direct consequence of
the Green-Schwarz term, and it is remarkable that the sign of $h_T$ is
indeed negative. Note that for $X^T$, $X^S$, $h_S$ and $h_T$ as given
in Eqs.~\eqref{XTS} and \eqref{hST}, $h > 0$ always implies $D > 0$
for $N \geq 1$. Hence, for nonzero flux, the matter fields are always
stabilized at the origin.

In the expression \eqref{simple} for the scalar potential only the term $-EB$ can become negative. For the 6d model considered here one has $z>0$ (see Eqs.~\eqref{zmoduli}, \eqref{XTS}). Hence, $B$ is bounded from below and from above, and $-EB$ can approach $-\infty$ only for $s,t \rightarrow 0$ (see Eq.~\eqref{Eparameter}). However, in this limit the positive third term in Eq.~\eqref{simple} is more singular. The scalar potential is therefore bounded from below.

We are now ready to apply the general analysis of the previous section
to our example. For simplicity, we set $\epsilon = 0$ in
Eqs.~\eqref{Ast} and look for Minkowski vacua. Obviously, there exist very similar solutions for de Sitter vacua with $\epsilon \gtrsim 0$. Having GUT scale extra
dimensions and three generations in mind \cite{Buchmuller:2015jna}, we choose
$\ell = 50$, $g = 0.2$ and $N = 3$. 
Using Eqs.~\eqref{Ast} and writing $az \equiv a_S (s + \kappa t)$,  
we find for a Minkowski minimum at $(s,t) \simeq (5.3, 9.4)$ typical
superpotential parameters,
\begin{align}
a_S \simeq 0.63 \, , \quad W_0 \simeq -9.1 \times 10^{-4} \, , \quad W_1 \simeq 6.9 \times 10^{-4} \,.
\label{parameters}
\end{align}
For the effective gauge coupling we obtain $g_\mathrm{eff} = g h^{-1/2} \simeq 0.16$, and the size of the
compact dimension is $V_2 = \ell^2\sqrt{st}/(2g^2)= (7\times
10^{15}\text{GeV})^{-2}$,
implying GUT scale supersymmetry breaking.
Note that also larger extra dimensions can be obtained without fine
tuning for larger values of $h$, i.e. for smaller gauge couplings and
smaller superpotential parameters $W_0$, $W_1$. The masses of all
moduli are of order $10^{14}\,\text{GeV}$. 
In Fig.\ \ref{minimum} the scalar potential is shown close to
the Minkowski minimum. It diverges as $s$ and $t$
approach the critical line $h=0$.
\begin{figure}[t]
\centering
\includegraphics[width=.65\textwidth]{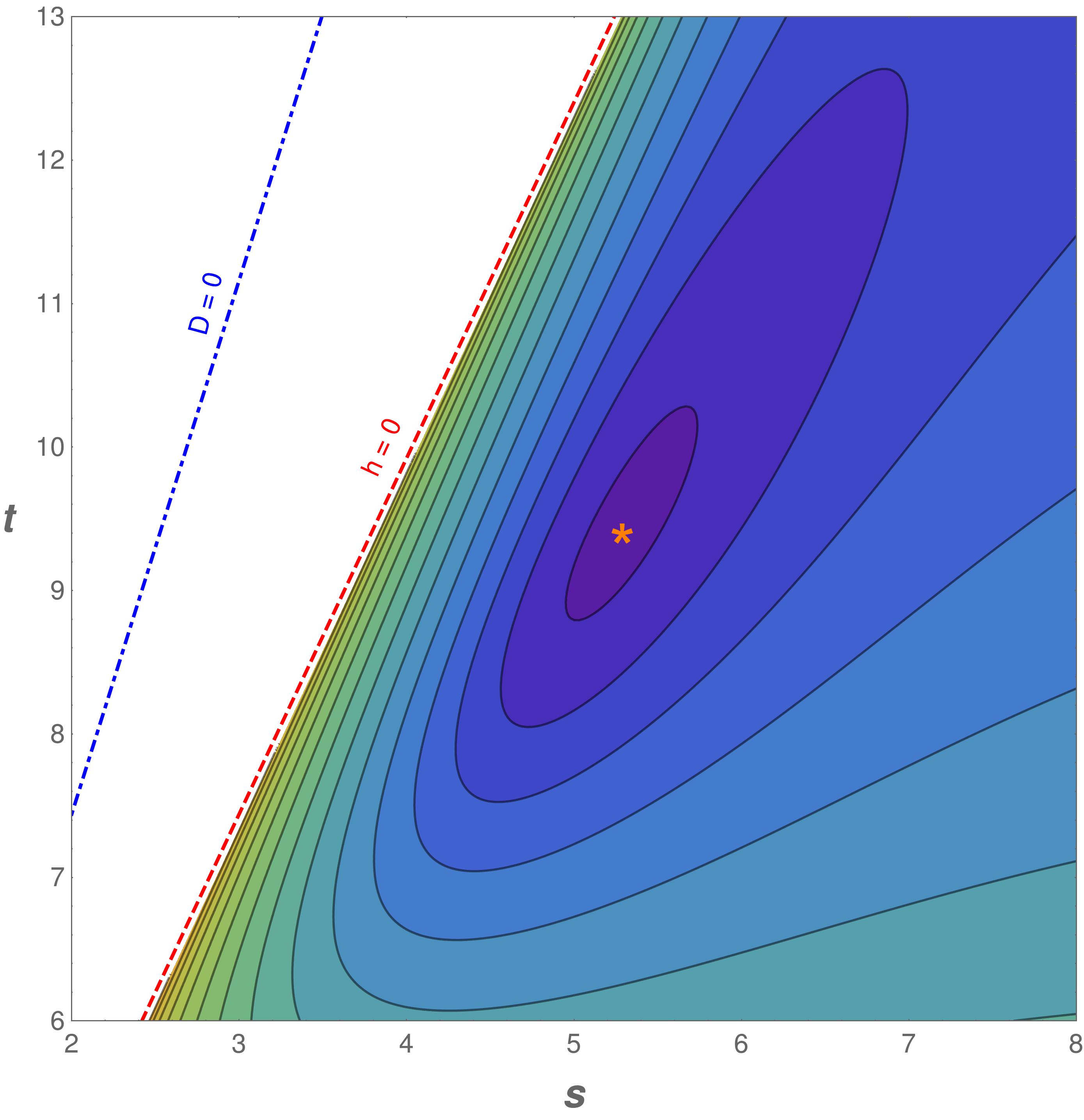}
\caption{Minkowski minimum for the superpotential parameters
  \eqref{parameters}. $h$ vanishes along the red dashed line \eqref{physicalregime} and $D$ vanishes along the
  blue, dashed-dotted line \eqref{Dzero}.
  The minimum is at $(s, t) = (5.3, 9.4)$ as indicated by
  the orange star.}
\label{minimum}
\end{figure}

\section{Discussion}
\label{sec:conclusion}

We have presented a new class of metastable de Sitter vacua. They arise
in no-scale supergravity models with a gauged shift symmetry involving
two moduli fields, $S$ and $T$. This leads to two Fayet-Iliopoulos
terms with different moduli dependence. Also the gauge kinetic function
has contributions linear in $S$ and $T$. The superpotential of KKLT-type depends on the gauge invariant linear combination of $S$ and
$T$. As we have shown, de Sitter or Minkowski vacua are naturally
obtained if the $T$-dependent contribution to the gauge kinetic function
is negative.

It is remarkable that the low-energy effective action of
six-dimensional supergravity with flux, compactified on an orbifold,
fulfills the sufficient conditions for de Sitter vacua. The
$T$-dependent FI term is generated by magnetic flux, whereas the
$S$-dependent FI term and the $T$-dependent contribution to the gauge
kinetic function are due to the Green-Schwarz term needed to
cancel the $U(1)$ gauge anomaly. The Green Schwarz-term leads
to the negative sign of the $T$-dependent part in the gauge kinetic function.

Flux and anomaly cancellation by the Green-Schwarz mechanism are generic
ingredients of string compactifications. We therefore believe that 
our discussion of de Sitter vacua is relevant far beyond the
six-dimensional supergravity example discussed in this Letter.
It will be interesting to study applications for the heterotic string
as well as for type IIB string theory and F-theory. Furthermore, it is
intriguing that typical values for the superpotential parameters and
the gauge coupling lead to GUT scale extra dimensions and a related large
supersymmetry breaking scale. This is a challenge for the electroweak
hierarchy problem, analogous to the cosmological constant problem
which is encoded in the fine tuning of the superpotential parameters.

\section*{Acknowledgments}
We thank Michele Cicoli, Emilian Dudas, Stefan Groot Nibbelink, Zygmunt Lalak, Jan Louis, Hans-Peter Nilles, Alexander Westphal and Clemens Wieck for valuable discussions. This work was supported by the German Science Foundation (DFG) within the Collaborative Research Center (SFB) 676 ``Particles,
Strings and the Early Universe''. M.D. acknowledges support from the Studienstiftung des deutschen Volkes.

}

\providecommand{\href}[2]{#2}\begingroup\raggedright\endgroup

\end{document}